\documentclass [12pt]{article}
\usepackage{epsf}

\catcode`@=12
\newcommand{\be}{\begin{equation}}
\newcommand{\ee}{\end{equation}}
\newcommand{\bea}{\begin{eqnarray}}
\newcommand{\eea}{\end{eqnarray}}

  \def\f{n_f}

\def\f12{\frac{1}{2}} 
\def\f16{\frac{1}{6}} 
\def\f23{\frac{2}{3}} 
\def\f13{\frac{1}{3}} 
\def\f43{\frac{4}{3}}

\begin{document}
\thispagestyle{empty}
\begin{flushright} 
SINP/TNP/02-14
\end{flushright}
\begin{center}
{\LARGE $E_6$ multiplets and unification in extra dimensions\\}
\vskip .5in
{\bf Biswajoy Brahmachari\footnote{Adress before August 2002:
Theoretical Physics Division, Saha Institute of Nuclear Physics,
Kolkata-64, India} \\}

\vskip 1cm
{\it
Department of Physics, 
Vidyasagar Evening College,\\
39, Sankar Ghosh Lane, Kolkata 700006, India \\
}

\end{center}

\vskip .5in

\begin{center}
\underbar{Abstract} \\
\end{center}
We study the effect of all matter multiplets contained in {\bf 27} 
representation of $E_6$ GUT on gauge coupling  unification in extra 
dimensions. Extra members of {\bf 27} multiplets of all three generations
have their `zero modes' near $m_t$ such that they can be directly probed.
From TeV scale onwards extra dimensions open up, theory becomes N=2 
supersymmetric and gauge couplings unify or they do not depending on how
we distribute matter fields and gauge fields in bulk and brane. We find 
three such possible embedding which will lead to perfect gauge coupling
unification below 100 TeV region for one extra dimension and lower than 
that if number of extra dimensions is larger. 

\newpage
Supersymmetry breaking is a hard problem. However attempts have been
made to link lightness of supersymmetry breaking scale to some large 
radius of compactifications in string theory\cite{antoniadis}.
Independently large extra dimensions of millimeter size have been invoked to
stabilize gauge hierarchy problem as proposed in Ref.\cite{arkani}.
This way to tackle hierarchy problem is independent of
the existence of low energy supersymmetry.
Moreover special features of open string theory\cite{witten} 
have also been used to try and bring down the fundamental string
scale itself to TeV region\cite{lykken,lowstring}. All these new results
give good motivation for studying theories where extra
dimensions show up much below $10^{18}$ GeVs. One possible
consequence of such large extra dimensions is that gauge coupling
unification can happen in $M_X=$ few tens of TeVs\cite{ddg}. Then even
though our unified theory in extra dimensions is non-renormalizable
gauge coupling unification happens close enough to the low lying string 
scale in such a way that three gauge forces unify with gravity and 
side by side divergences of non-renormalizable effective gauge theory 
are properly handled by full string theory which takes over almost 
immediately.

Here we will study gauge coupling unification in $E_6$ 
Grand Unified Theory (GUT)\cite{gur} in
extra dimensions. We know that fermions are unified in
the {\bf 27} representation of $E_6$ which can be decomposed
as
\bea
E_6 \supset && SU(3) \times SU(2) \times U(1)  \nonumber \\
27   \supset &&  \overbrace{(3,2,1/6)+(\overline{3},1,-2/3)+(1,1,1)}
^{10~of~SU(5)}
+\overbrace{(\overline{3},1,1/3) +(1,2,-1/2)}^{\overline{5}~of~SU(5)}
+\overbrace{(1,1,0)}^{RH~neutrino} \nonumber\\
&&\underbrace{ +(\overline{3},1,1/3)+(3,1,-1/3)
+(1,2,-1/2)+(1,2,1/2)+(1,1,0)}_{exotic~fermions}. \nonumber
\eea
Because two extra singlets will not affect gauge coupling unification
we will work with low energy multiplets that can be thought
as $5+\overline{5}$ of SU(5). We know that in 4 dimensions 
introduction of complete SU(5) multiplets do not affect
gauge coupling unification but the unified coupling
can become non-perturbative before GUT scale is reached. However as we 
will learn below this is not necessarily the case in the presence of 
extra dimensions. Because this is a three generation analysis 
we can have full three copies of $5 +\overline{5}$ hanging well below 
the GUT scale due to $E_6$ symmetry. That is $5 +\overline{5}$ do not 
pair up and become as heavy as the GUT scale. Then we have four possible 
types of extra matter multiplets. We assume that their masses are near 
the top quark mass $m_t$. In extra dimensional models we use the terminology
that their `zero modes' are near $m_t$. At the scale $\mu_0= 1$ TeV 
$\delta$ number of extra dimensions open up and excited
Kaluza-Klein states starts to show. Under standard model 
gauge group we label extra matter superfields as,
\be
L_1=(1,2,-1/2)~~;~~ D_1=(\overline{3},1,1/3)
~~;~~ L_2=(1,2,1/2)~~;~~ D_2=(3,1,-1/3).
\ee
Contribution of $L_1$ and $L_2$ to beta coefficients will be same as usual 
lepton doublet whereas contribution of $D_1$ and $D_2$ will be same as down 
type antiquark.
\begin{table}[htb]
\begin{center}
\[
\begin{array}{cc|c|c|c}
\hline
fields & representation & \tilde{\beta}_3 & \tilde{\beta}_2 
& \tilde{\beta}_1\\
\hline
H_1 & (1,2, {1 \over 2})  & 0 & 1 & 1\\
H_2 & (1,2, -{1 \over 2})  & 0 & 1& 1\\
Q & (3,2, -{1 \over 6})  & \eta_Q~2& 
\eta_Q~3& \eta_Q~{1 \over 3}\\  
\overline{D} & (\overline{3},1, {1 \over 3})  & 
\eta_U~1& 
0 & 
\eta_U~{2 \over 3}\\  
\overline{U} & (\overline{3},1, -{2 \over 3})  & 
\eta_D~1 & 
0 & 
\eta_D~{8 \over 3}\\  
L & (1,2, -{1 \over 2})  & 
0 & 
\eta_L~1& 
\eta_L~1\\  
\overline{E} & (1,1, 1)  & 
0 & 
0 & 
\eta_E~2\\  
\hline
gauge & (8,3, 0)  & 
-6 & 
-4& 
0\\  
\hline
\end{array}
\]
\end{center}
\caption{
contributions to $\tilde{\beta}$ coefficients in (N=2) supersymmetric
standard model. Extra $E_6$ multiplets $L_1$ and
$L_2$ contribute same as L and $D_1$ and $D_2$ contribute same 
as $\overline{D}$} 
\label{table2}
\end{table}
Gauge couplings evolve\cite{ddg} with energy via the following equation where
we have redefined 
$t={1 \over 2 \pi} {\rm ln} (\Lambda)$,
$t_0={1 \over 2 \pi} {\rm ln} (\mu_0)$ and
$\alpha=g^2/4 \pi$. 
Here $\Lambda$ is the cut-off scale where couplings are being evaluated.
\be
{ d \alpha_i \over dt}= [(\beta_i - \tilde{\beta}_i) + 
\tilde{\beta}_i ~X_\delta ~e^{2 ~\pi ~\delta~(t-t_0)}]~ \alpha^2_i~~~
{\rm and}~~~ X_\delta={ \pi^{\delta/2} \over \Gamma(1+\delta/2)}. \label{rge}
\ee
$\Gamma$ is Euler Gamma function. Note that we must recover familiar 
Renormalization Group Equations (RGE)
in the limit of either $\tilde{\beta}=0$ or $\delta=0$. We have used
the usual notation\cite{papers} that $\beta$s are 4-dimensional 
coefficients whereas $\tilde{\beta}$s are higher dimensional
coefficients. Expressions of $\tilde{\beta}$s are given in detail
in Tables (1) for N=2 supersymmetric standard model.

\begin{figure}
\begin{tabular}{cc}
\epsfysize=8cm \epsfxsize=8cm \hfil \epsfbox{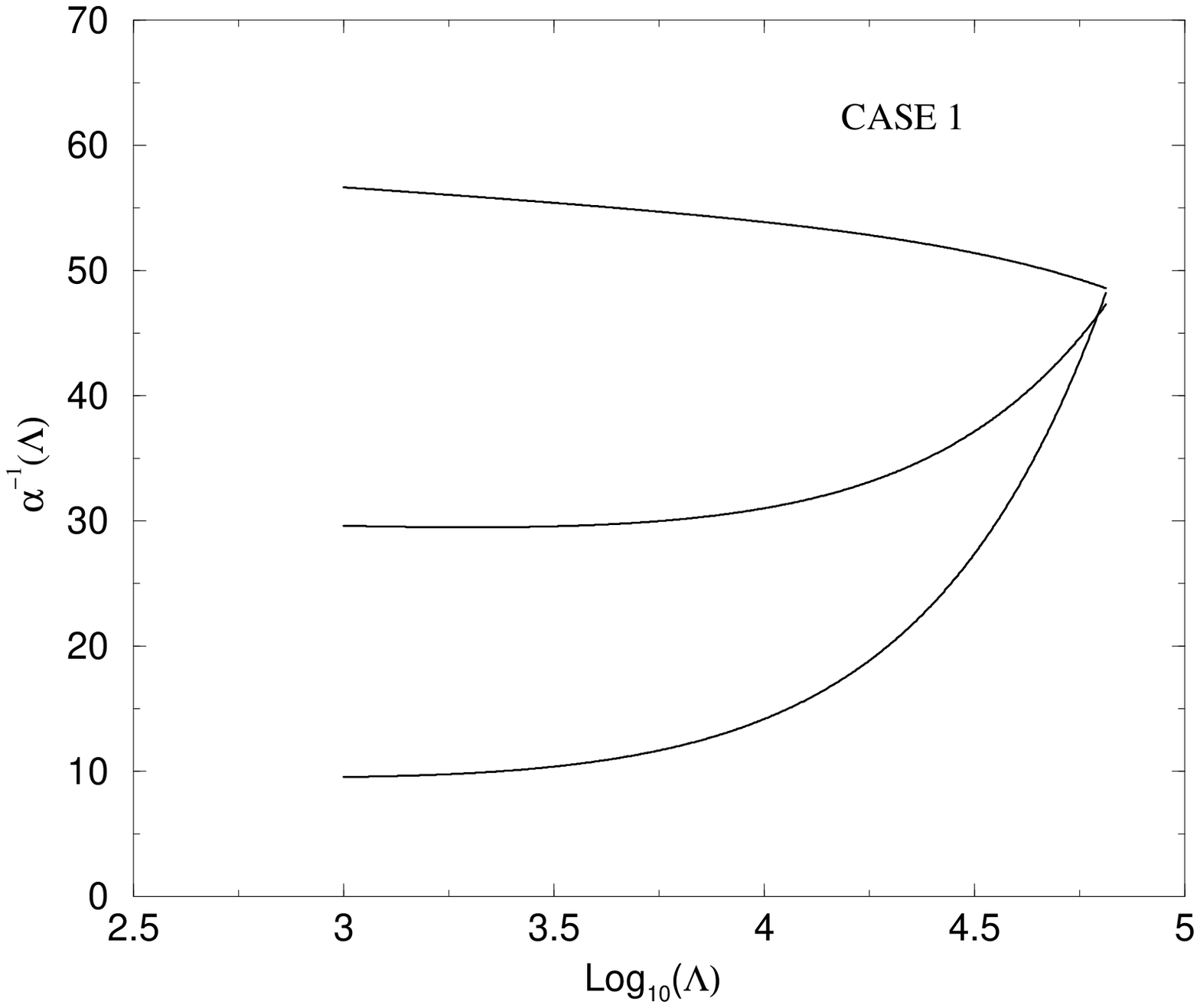} \hfil
& \epsfysize=8cm \epsfxsize=8cm \hfil \epsfbox{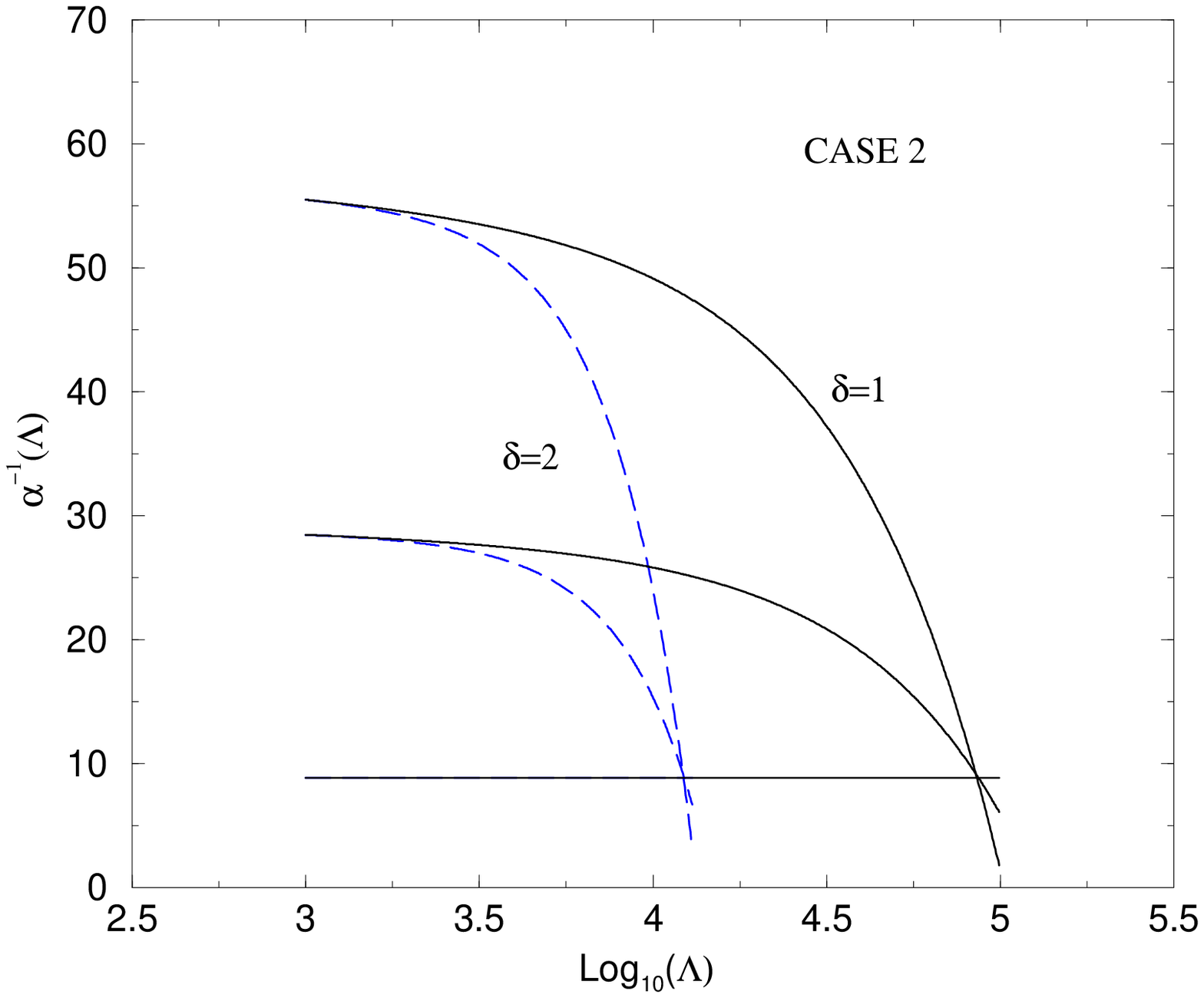} \hfil
\\ 
\epsfysize=8cm \epsfxsize=8cm \hfil \epsfbox{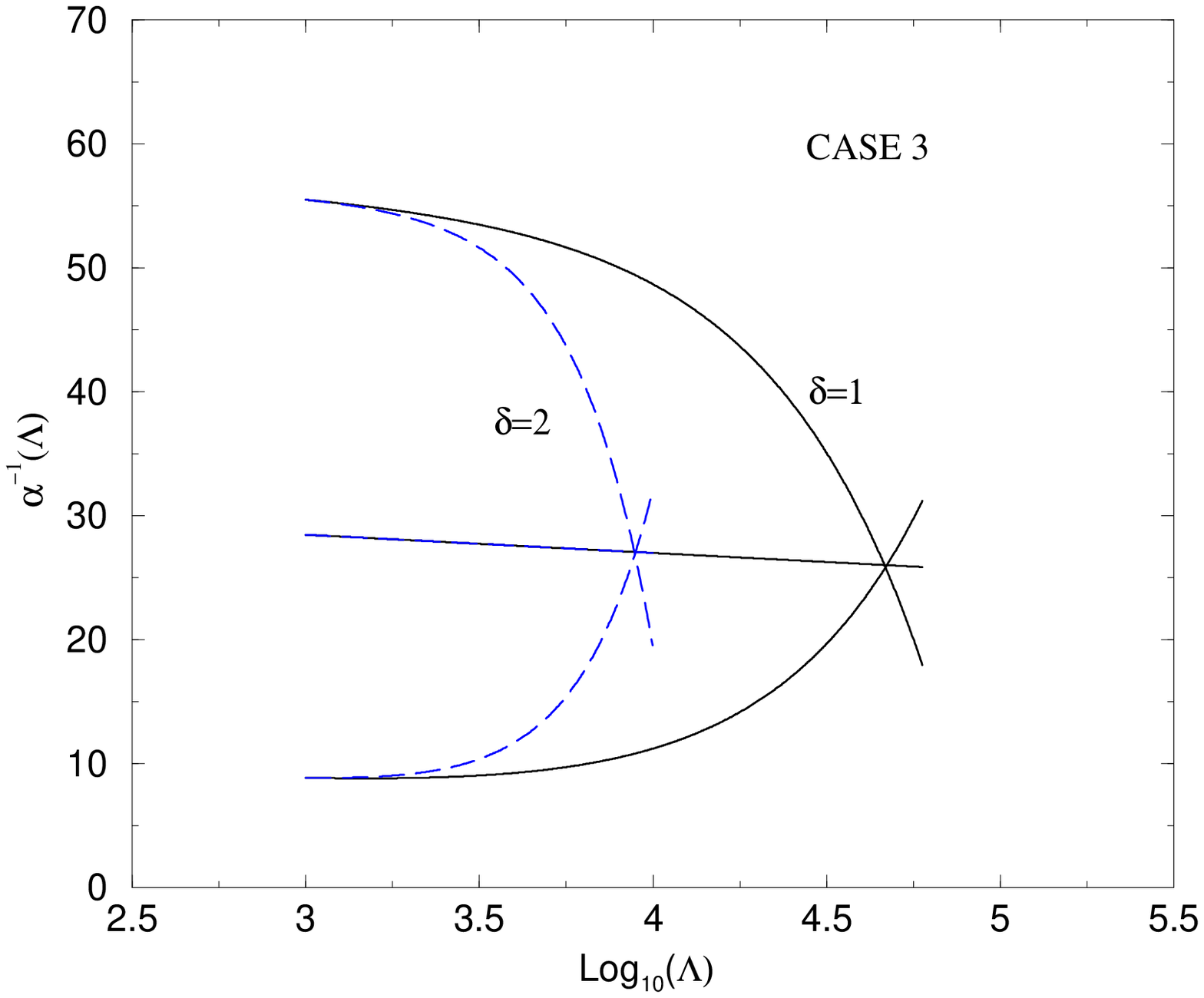} \hfil
& \epsfysize=8cm \epsfxsize=8cm \hfil \epsfbox{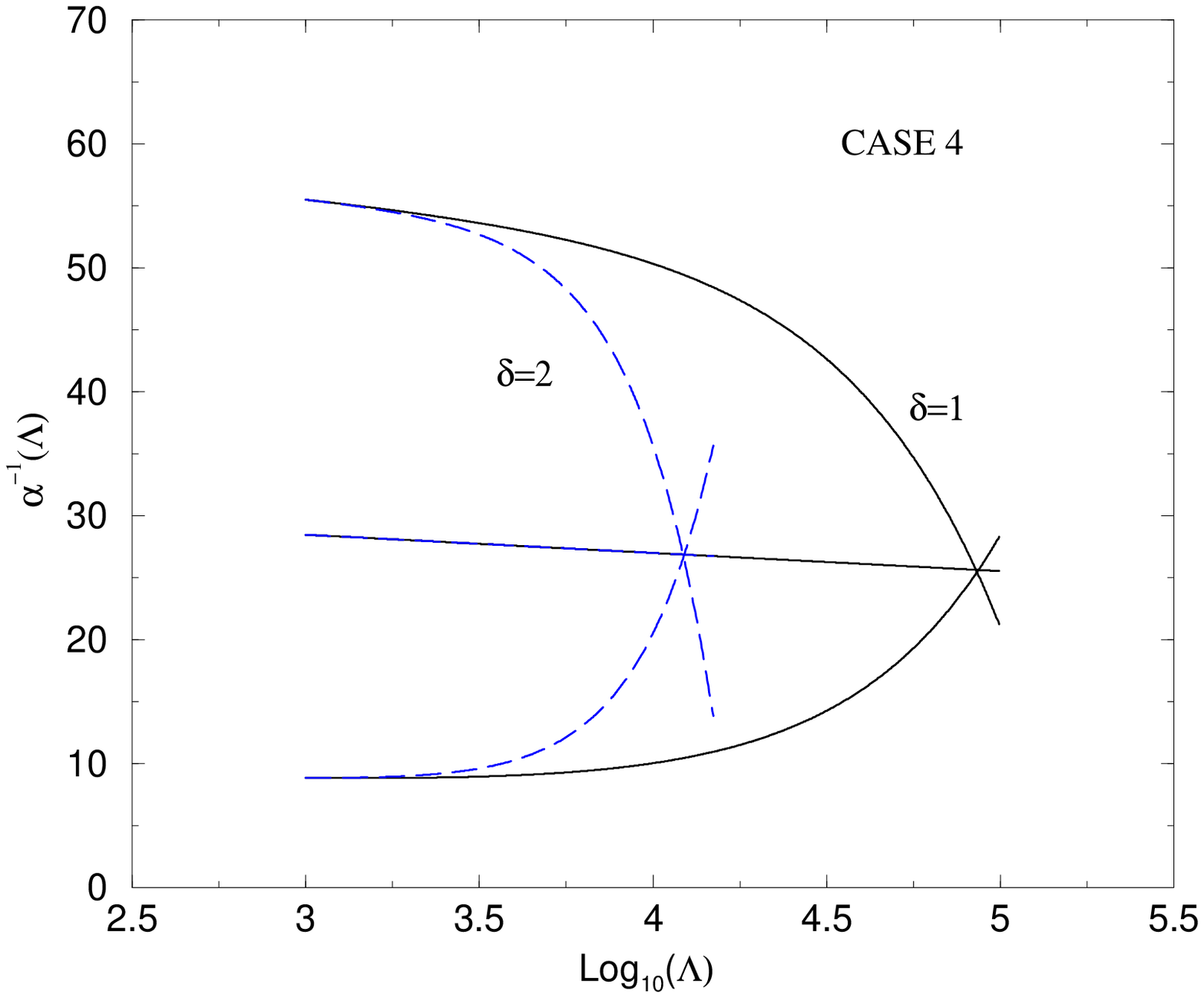} \hfil
\end{tabular}
\caption{CASE 1: 
Original DDG case with $\mu_0 \sim 1$ TeV where only gauge bosons and
Higgs scalar has bulk excitations. Gauge couplings do not meet
precisely.
CASE 2:
Only SU(2) and U(1) has bulk excitations.
Three {\bf 27} generations at scale $m_t$ of which leptons have
bulk excitations with $\eta_L=5,\eta_E=1$ above $\mu_0$.
CASE 3:
Only SU(3) and U(1) has bulk excitations. All three 27 multiplets have 
zero modes at scale $m_t$, above $\mu_0$ $\eta_E=2,\eta_U=2$ have
bulk excitations.
CASE 4:
Only SU(3) and U(1) has bulk excitations. All three 27 multiplets have 
zero modes at scale $m_t$, above $\mu_0$ $\eta_D=3, \eta_U=2$ have
bulk excitations.
 }
\label{fig1}
\end{figure}

Let us quickly review the minimal scenario given by Dines Dudas and
Gherghetta\cite{ddg} which will set all our notations.
There only gauge bosons and Higgs
have bulk excitations whereas fermions remain at fixed
points. Two Higgs doublets $H_1$ and $H_2$ are embedded in a single
$N=2$ Higgs superfield. All $\delta$ extra dimensions open up 
simultaneously at scale $\mu_0$ which is a free parameter.
Below the scale $\mu_0$ we have N=1 supersymmetry whereas
above $\mu_0$ we have N=2 supersymetry. In this case 
$\beta=({33 \over 5},1,-3)$ and
$\tilde{\beta}=({3 \over 5},-3,-6)$. Following Eqn.(\ref{rge}) we
write
\bea
{d \alpha_Y \over dt}
&=&[6+3/5~X_\delta ~e^{2 ~\pi ~\delta~(t-t_0)}]~\alpha^2_Y, \nonumber\\
{d \alpha_2 \over dt}
&=&[4-3~X_\delta ~e^{2 ~\pi ~\delta~(t-t_0)}]~\alpha^2_2, \nonumber\\
{d \alpha_3 \over dt}
&=&[3-6~X_\delta ~e^{2 ~\pi ~\delta~(t-t_0)}]~\alpha^2_3. \label{ddgeq}
\eea
Using two loop renormalization group equations for the
gauge couplings below $\mu_0$ we can solve Eqn. (\ref{ddgeq})
numerically, the results are given in Fig. (\ref{fig1}) CASE 1.
Now let us examine the unification condition discussed by
DDG by defining ratio
\be
R_{ij}={\tilde{\beta}_i -\tilde{\beta}_j \over \beta_i -\beta_j}.
\label{ratio}
\ee
Then unification is achieved when we have,
\be
R_{12}=R_{13}=R_{23}\label{cond}.
\ee
For DDG case we have ${R_{12} \over R_{13}}=0.94$ and 
${R_{13} \over R_{23}}=0.92$. Thus gauge coupling unification
is only approximate. Let us examine scenario 3 discussed
by Carone\cite{carone}. There are two $5 +\overline{5}$ of which
leptons have bulk excitations (total 5 N=1 pairs) and one generation
of electron have bulk excitation (1 N=1 pair). All fermions have
zero modes at scale O($m_W$). SU(3) gauge bosons stay on
the boundary. Then we have $R_{12}=R_{13}=R_{23}=1/2$ which gives 
perfect unification. We see this case is very similar to our 
$E_6$ scenario except that we want to keep zero modes of
all three generations of {\bf 27} near $m_t$. 
This fixes $\beta$ coefficients below $\mu_0$ to be
$\beta=(48/5,4,0)$. However adding one more $5 + \overline{5}$ 
near $m_t$ will keep the difference $\beta_i-\beta_j$ unchanged
which occurs in the denominator in Eqn. (\ref{ratio}). Thus couplings 
will still unify. Keeping only SU(2) and U(1) gauge
bosons in the bulk we calculate unification condition using Eqn. (\ref{cond}),
\be
\eta_E= {3 \eta_L - 13 \over 2}.
\ee
For $\eta_L=5,\eta_E=1$ we get CASE 2
\be
\beta=(48/5,4,0)~~~,~~~
\tilde{\beta}=(24/5,2,0)
\ee
Then three generations of full {\bf 27} multiplets contribute
to evolution of gauge couplings from the scale $m_t$ onwards
and changes coefficient $\beta$ even though their effect does
not show up in the difference $\beta_i - \beta_j$ which
appears in the quantity $R_{ij}$.

Next let us consider the case when above $\mu_0$ SU(3) and U(1) gauge 
bosons have bulk excitations but SU(2) do not.
Again because bulk matter transform under bulk gauge group
only $\overline{U},\overline{D}$ and $\overline{E}$ can have bulk
excitations. Then we calculate unification condition using Eqn. (\ref{cond}) 
\be
\eta_D={ 14 - 2 \eta_E - 5 \eta_U \over 3}.
\ee
and two pairs of $\overline{E}$ and two pairs of
$\overline{U}$ can have bulk excitations. Then below and above
$\mu_0$ the beta coefficients
are
\be
\beta=(48/5,4,0)~~~,~~~\tilde{\beta}=(28/5,0,-4).
\ee
In this scenario we get,
\be
R_{12}=R_{13}=R_{23}=1.
\ee
The unification picture is given in Fig (\ref{fig1}) CASE 3. Second
case is when three $\overline{D}$ pairs and one 
$\overline{U}$ pair have bulk excitations.
In this case we get,
\be
\beta=(48/5,4,0)~~~,~~~\tilde{\beta}=(14/5,0,-2).
\ee
In this scenario we get,
\be
R_{12}=R_{13}=R_{23}=1/2.
\ee
The unification picture is given in Fig (\ref{fig1}) CASE 4.

Even though strictly speaking Eqn(\ref{cond}) is valid at one loop,
below $\mu_0$ we have used two loop running assuming all superpartners
and extra $E_6$ matter near the scale $m_t$. This does not affect gauge
coupling unification appreciably as we see from Fig (\ref{fig1}).
Also form Fig (\ref{fig1}) we see that for $\delta=2$ unification
scale is much lower than $\delta=1$ case. 
We would like to stress that we have not introduced any new
multiplet in adhoc basis. We have only concentrated on multiplets
contained within {\bf 27} representation of $E_6$ group.
In next generation colliders 
`zero modes' of these the $E_6$ exotic particles may be 
discovered\cite{colle6}. Furthermore excited Kaluza-Klein modes of 
gauge bosons and matter fields will also be rigorously 
searched \cite{collkk} in near future. 

We have ignored heavy threshold corrections in the paper. This
is because we have assumed that heavy GUT multiplets exist at
or above the unification scale and they are degenerate in
mass. This is in the spirit of an extended survival hypothesis in
convensional unified models. However in case some odd members of a 
heavy GUT multiplets have masses below the unification scale we will 
need to include heavy threshold corrections\cite{hth} to our results.

In a recently updated study Kang and Langacker have 
studied the discovery limits\cite{lang1} of exotic $E_6$ multiplets in
Fermilab Tevatron and CERN LHC. They conclude that multiplets
as light as 200 GeV can be probed directly in Tevatron
and as light as 1 TeV in LHC. A natural question in this context 
is that are we allowed to put exotic $E_6$ multiplets will masses 
near $m_t$? In another words are they excluded by direct and indirect 
experimental searches? The answer is that with present experimental accuracy
exotic $E_6$ multiplets are allowed at around the scale
$m_t$\cite{lowerlimit}.

I would like 
to thank M. S. Berger for communications on Eqn (\ref{rge}). {\tt 
COPYRIGHT HOLDER OF THIS ARTICLE IS ACTA PHYSICA POLONICA
B}

\end{document}